\documentclass[10pt,conference]{IEEEtran}

\AtBeginDocument{%
  \providecommand\BibTeX{{%
    Bib\TeX}}}
    
\IEEEoverridecommandlockouts

\usepackage{xspace}
\usepackage{pifont}
\usepackage{xcolor}
\usepackage{multirow}
\usepackage{enumitem}
\usepackage{url}
\usepackage{booktabs}

\usepackage{subcaption}
\usepackage{tcolorbox}
\usepackage[linesnumbered,ruled,vlined]{algorithm2e}
\usepackage[table,xcdraw]{xcolor}

\usepackage{cite}
\usepackage{amsmath,amssymb,amsfonts}
\usepackage{graphicx}
\usepackage{textcomp}
\usepackage{wasysym}

\def\eg{\emph{e.g.,}\xspace} 

\def\ie{\emph{i.e.,}\xspace} 

\def\method{{\sc CoLA}\xspace}
\def\BibTeX{{\rm B\kern-.05em{\sc i\kern-.025em b}\kern-.08em
    T\kern-.1667em\lower.7ex\hbox{E}\kern-.125emX}}

\newcommand{\changeline}[1]{\textcolor{black}{#1}}
\newenvironment{change}[0]{\par\color{black}}{\par}
\newcommand{\changeref}[0]{\color{black}}

\begin{document}

\title{Aligning LLMs to Fully Utilize the Cross-file Context in Repository-level Code Completion}

\author{
\IEEEauthorblockN{Jia Li \male$^{*}$ \thanks{* Equal contribution. $\dagger$ Corresponding author.}}
\IEEEauthorblockA{
\textit{Peking University} \\
Beijing, China \\
jia\_li@mail.tsinghua.edu.cn}
\and
\IEEEauthorblockN{Hao Zhu$^{*}$}
\IEEEauthorblockA{
\textit{Peking University} \\
Beijing, China \\
zhuhao@stu.pku.edu.cn}
\and
\IEEEauthorblockN{Huanyu Liu$^{*}$}
\IEEEauthorblockA{
\textit{Peking University} \\
Beijing, China \\
huanyuliu@stu.pku.edu.cn}
\and
\IEEEauthorblockN{Xianjie Shi}
\IEEEauthorblockA{
\textit{Peking University} \\
Beijing, China \\
2100013180@stu.pku.edu.cn}
\and
\IEEEauthorblockN{He Zong}
\IEEEauthorblockA{
\textit{aiXcoder} \\
Beijing, China \\
zonghe@aixcoder.com}
\and
\IEEEauthorblockN{Yihong Dong, Kechi Zhang}
\IEEEauthorblockA{
\textit{Peking University} \\
Beijing, China \\
\{dongyh, zhangkechi\}@[stu.]pku.edu.cn}
\and
\IEEEauthorblockN{Siyuan Jiang}
\IEEEauthorblockA{
\textit{aiXcoder} \\
Beijing, China \\
jiangsiyuan@aixcoder.com}
\and
\IEEEauthorblockN{Zhi Jin, Ge Li$^{\dagger}$}
\IEEEauthorblockA{
\textit{Peking University} \\
Beijing, China \\
\{zhijin, lige\}@pku.edu.cn}
}
\maketitle


\begin{abstract}
Large Language Models (LLMs) have shown promising results in repository-level code completion, which completes code based on the in-file and cross-file context of a repository.
The cross-file context typically contains different types of information (\eg relevant APIs and similar code) and is lengthy. In this paper, we found that LLMs struggle to fully utilize the information in the cross-file context. We hypothesize that one of the root causes of the limitation is the misalignment between pre-training (\ie relying on nearby context) and repo-level code completion (\ie frequently attending to long-range cross-file context). 

To address the above misalignment, we propose Code Long-context Alignment - \method, a purely data-driven approach to explicitly teach LLMs to focus on the cross-file context. Specifically, \method constructs a large-scale repo-level code completion dataset - \method-132K, where each sample contains the long cross-file context (up to 128K tokens) and requires generating context-aware code (\ie cross-file API invocations and code spans similar to cross-file context). Through a two-stage training pipeline upon \method-132K, LLMs \changeline{learn} the capability of finding relevant information in the cross-file context, thus aligning LLMs with repo-level code completion. We apply \method to multiple popular LLMs (\eg aiXcoder-7B) and extensive experiments on \method-132K and a public benchmark - CrossCodeEval. Our experiments yield the following results. \ding{182} \textit{Effectiveness.} \method substantially improves the performance of multiple LLMs in repo-level code completion. For example, it improves aiXcoder-7B by up to 19.7\% in exact match. \ding{183} \textit{Generalizability.} The capability learned by \method can generalize to new languages (\ie languages not in training data). \ding{184} \textit{Enhanced Context Utilization Capability.} We design two probing experiments, which show \method improves the capability of LLMs in utilizing the information (\ie relevant APIs and similar code) in cross-file context. Our datasets and model weights are released in \cite{Replication_Package}.
\end{abstract}


\section{Introduction}
\label{sec:introduction}

Repository-level (Repo-level) code completion aims to complete an unfinished code file based on the in-file context and cross-file context within the current repository \cite{RepoCoder,GraphCoder,CoCoMIC}. The cross-file context typically contains different types of information across files (\eg relevant APIs \cite{CoCoMIC} and similar code \cite{RepoCoder,GraphCoder}) and is lengthy (up to thousands of tokens). In recent years, Large Language Models (LLMs) have achieved State-Of-The-Art (SOTA) performance in repo-level code completion \cite{aixcoder,DeepSeek-Coder,CodeLlama}. To process the long cross-file context, the context windows of LLMs are often large, \eg 16,384 tokens.

However, we found that existing LLMs struggle to fully utilize information within the cross-file context in repo-level code completion. Our motivation stems from two observations. First, recent studies \cite{Lost_in_the_Middle1,Lost_in_the_Middle2} revealed that LLMs fail to pass simple probing tasks such as Needle-in-the-Haystack, which asks LLMs to recall specific text segments in long contexts. Second, we conducted a preliminary experiment and found similar problems in repo-level code completion. For example, LLMs fail to utilize relevant APIs and similar code presented in the cross-file context. Section \ref{sec:motivating_ex} shows two examples and detailed analyses. However, recent studies \cite{RepoCoder,GraphCoder,RLCoder} focus on how to extract relevant context and ignore this severe limitation. Consequently, a pressing research question arises: \textit{How to make LLMs fully utilize the information in the cross-file context?}

We hypothesize that one of the root causes of the above limitation is the misalignment between pre-training and repo-level code completion. The pre-training data typically contains extensive files (\eg over 300 million files \cite{StarCoder}) from different sources, and the correlation between files is often sparse. In auto-regressive pre-training, the loss on predicting the next token is more likely to be influenced by a few nearby pre-tokens rather than long-distance tokens \cite{Pre-training_lost1,Pre-training_lost2}. This training may lead LLMs to focus mainly on the nearby context and ignore the long-distance context. In contrast, in repo-level code completion, the cross-file context is typically lengthy, and LLMs need to frequently attend to the long-distance context. As a result, the misalignment between pre-training and repo-level code completion limits the capabilities of LLMs in utilizing the cross-file context.

Based on this hypothesis, this paper proposes \textbf{Co}de \textbf{L}ong-context \textbf{A}lignment - \method, which explicitly teaches LLMs that the crucial information can be presented in the cross-file context, not just in the nearby context. \method is a purely data-driven solution that constructs a repo-level code completion dataset - \method-132K. Each sample in the dataset is equipped with the long cross-file context (up to 128K tokens), and asks LLMs to generate context-aware code. In this paper, we focus on two types of crucial information (\ie relevant APIs and similar code) in the cross-file context, which are widely used in related works \cite{RepoCoder,GraphCoder,RLCoder}. Thus, \method asks LLMs to generate two types of context-aware code, \ie cross-file API invocations and code spans similar to cross-file context. Intuitively, to predict both types of code, LLMs need to master the capability of finding relevant information in the cross-file context, thus aligning LLMs with repo-level code completion.

To achieve the above goal, we construct \method-132K from 2,000 high-quality open-source repositories, obtaining 120,000 training samples and 12,000 testing samples. Besides, \method-132K covers four popular programming languages (\ie Python, Java, C++, and Go). We further propose a two-stage pipeline to train LLMs on \method-132K. The training pipeline consists of a supervised fine-tuning stage and a reinforcement learning stage. More details are in Section \ref{sec:CoLT}.

To highlight the effectiveness of \method, we apply \method to multiple popular LLMs, including aiXcoder-7B \cite{aixcoder}, DeepSeek-Coder-6.7B \cite{DeepSeek-Coder}, and Code Llama-7B \cite{CodeLlama}. The model trained from aiXcoder-7B is named aiXcoder-7B-v2. The training details are in Section \ref{sec:aiXcoder}. 

To evaluate \method, we conduct a large-scale study on \method-132K (test set) and a public benchmark - CrossCodeEval \cite{CrossCodeEval}. Evaluation metrics are Exact Match (EM) and BLEU \cite{CodeBleu}. Our study yields the following results. 
\ding{182} \textit{Effectiveness.} \method substantially improves the performance of multiple LLMs in repo-level code completion. For example, it improves aiXcoder-7B by up to 19.7\% in EM. aiXcoder-7B-v2 even surpasses larger models (\eg DeepSeek-Coder-33B). We also exclude the impact of additional training data. 
\ding{183} \textit{Generalizability.} From the perspective of languages, the capability learned by \method can generalize to new languages (\ie languages not in \method-132K), such as C\# and TypeScript. From the perspective of models, \method is model-agnostic and effectively improves multiple LLMs, \eg DeepSeek-Coder and Code Llama. 
\ding{184} \textit{Enhanced Context Utilization Capability.} \method significantly improves the capability of LLMs in utilizing two types of information (\eg APIs and similar code) within the cross-file contexts. For example, in more than 90\% of test samples, aiXcoder-7B-v2 can accurately locate and invoke the relevant APIs. \ding{185} \textit{Ablation Study.} The results of the ablation study show that two training stages and two types of context-aware code contribute to \method.

The key contributions of this paper are listed as follows:

\begin{itemize}
    \item We find that LLMs struggle to fully utilize the cross-file context in repo-level code completion. To address this limitation, we propose \textbf{Co}de \textbf{L}ong-context \textbf{A}lignment - \method, which is a data-driven solution and explicitly trains LLMs to find relevant information in the cross-file context.
    \item To support \method, we collect and release a repo-level code completion dataset - \method-132K. It asks LLMs to generate context-aware code (\ie cross-file API invocations and code spans similar to the cross-file context).
    \item We apply \method to multiple popular LLMs and conduct extensive experiments. Results show that \method substantially improves the performance of LLMs on repo-level code completion.
\end{itemize}

\section{Motivating Examples}
\label{sec:motivating_ex}

In this section, we conduct a case study to show our motivation, \ie existing LLMs struggle to fully utilize the information within the cross-file context.

\begin{figure}[t]
\centering
\includegraphics[width=\linewidth]{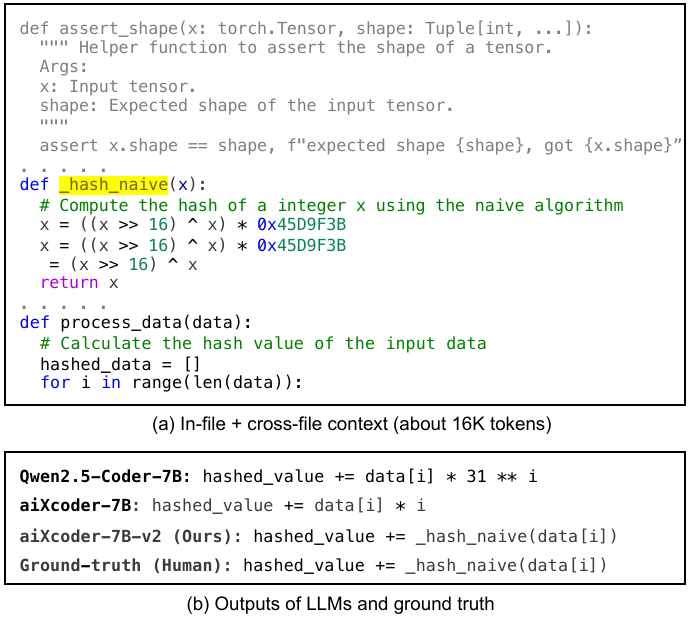}
\caption{Motivating Example \ding{182}. LLMs fail to invoke a relevant API in the cross-file context, leading to suboptimal completions.}
\label{fig:motivating_ex1}
\vspace{-0.6cm} 
\end{figure}

\textbf{Motivating Example \ding{182}: LLMs fail to invoke the relevant API in the cross-file context, leading to suboptimal results.} Figure \ref{fig:motivating_ex1} shows a real-world example of repo-level code completion. In this example, LLMs need to complete an unfinished Python function - \texttt{process\_data}, which calculates the hash value of input data. There is a relevant API - \texttt{\_hash\_naive} in the cross-file context, which provides a developer-customized approach to computing the hash value. Figure \ref{fig:motivating_ex1} (b) shows the outputs of existing LLMs and the ground-truth code. Qwen2.5-Coder-7B \cite{Qwen2.5-Coder} and aiXcoder-7B \cite{aixcoder} are two popular LLMs for code and achieve SOTA results in code completion. We can see that human developers invoke the API - \texttt{\_hash\_naive} to compute hash values. However, Qwen2.5-Coder-7B and aiXcoder-7B struggle to invoke the relevant API and output suboptimal completions.

\begin{figure}[t]
\centering
\includegraphics[width=\linewidth]{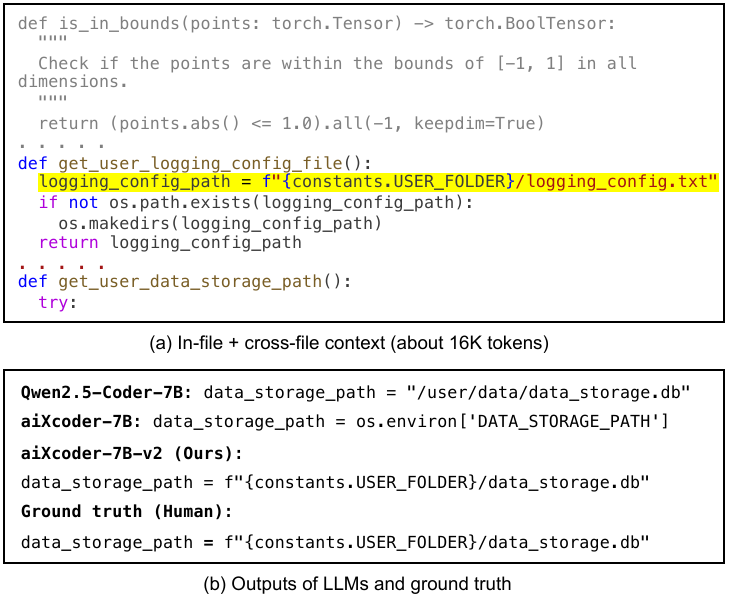}
\caption{Motivating Example \ding{183}. LLMs struggle to utilize similar code in the cross-file context, outputting wrong results.}
\label{fig:motivating_ex2}
\vspace{-0.6cm} 
\end{figure}

\textbf{Motivating Example \ding{183}: LLMs struggle to utilize similar code in the cross-file context, outputting wrong results.}  Figure \ref{fig:motivating_ex2} shows a real-world example of repo-level code completion. LLMs require completing an unfinished function - \texttt{get\_user\_data\_storage\_path}, which returns the path of the user’s data storage. The cross-file context contains a function - \texttt{get\_user\_logging\_config\_path} with a similar purpose, which returns the path of the user’s logging config file. The similar function shows how to access the user’s folder (\ie \texttt{constants.USER\_FOLDER}) and is helpful to complete code. Figure \ref{fig:motivating_ex2} (b) shows the outputs of existing LLMs and the ground-truth code. Human developers refer to the similar function and write the correct path to the user’s data storage. However, both LLMs do not utilize the information in the similar function and output wrong completions.

The above two examples show that existing LLMs struggle to fully utilize the cross-file context in repo-level code completion. Recent studies \cite{Lost_in_the_Middle1,Lost_in_the_Middle2} also found similar phenomena in natural language understanding tasks.

As stated in Section \ref{sec:introduction}, we hypothesize that one root cause of the above limitation is the misalignment between pre-training and repo-level code completion. To address this misalignment, our idea is to train LLMs to generate context-aware code and push LLMs to actively utilize the information in the cross-file context. For example, similar to Figure \ref{fig:motivating_ex1}, we train LLMs to predict a cross-file API invocation. To minimize the generation loss, LLMs must learn to find relevant APIs from the lengthy cross-file context, thus naturally aligning to repo-level code completion. Similarly, we train LLMs to generate code similar to the cross-file context and teach LLMs to refer to similar code in the cross-file context. We further propose a two-stage training pipeline to implement the alignment. We apply \method to a SOTA LLM - aiXcoder-7B and present aiXcoder-7B-v2. Figure \ref{fig:motivating_ex1} \& \ref{fig:motivating_ex2} (b) show the outputs of aiXcoder-7B-v2. aiXcoder-7B-v2 successfully utilizes the information in the cross-file context and outputs correct completions.

\section{\method}
\label{sec:CoLT}

In this section, we propose \textbf{Co}de \textbf{L}ong-context \textbf{A}lignment - \method for aligning LLMs to repo-level code completion. Because \method is purely a data-driven solution, we first present a dataset - \method-132K and describe a two-stage training pipeline.

\subsection{Dataset - \method-132K} 
\label{sec:CoLT:data}

The idea of \method is to train LLMs to generate context-aware code and push LLMs to utilize information in the cross-file context. Considering existing repo-level code completion datasets (\eg RepoEval \cite{RepoCoder}) are small-scale and lack context-aware code, we construct a new dataset  - \method-132K. 

\subsubsection{Data Statistics}
\label{sec:CoLT:data:statistics}

\begin{figure}[t]
\centering
\includegraphics[width=\linewidth]{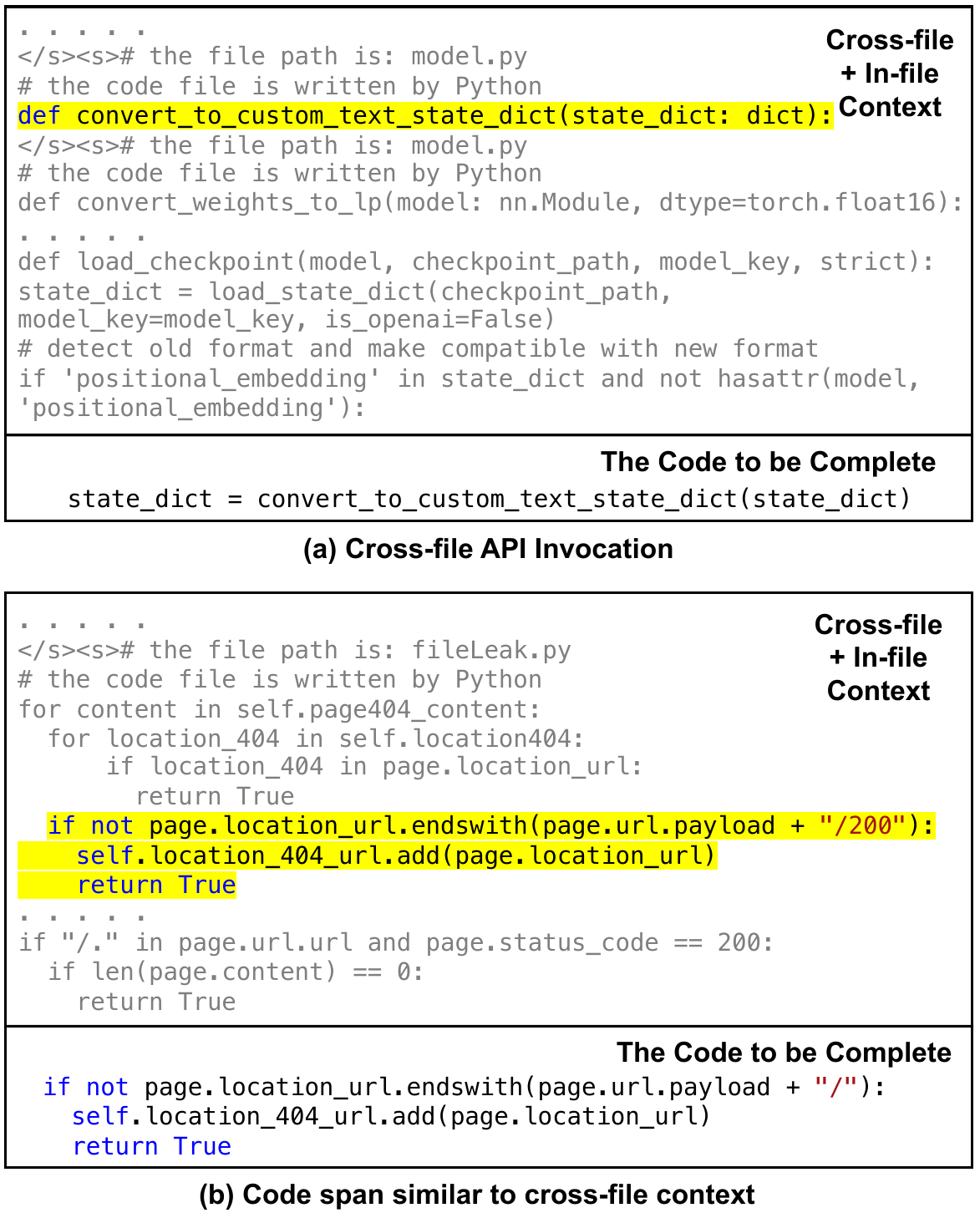}
\caption{Two samples in \method-132K.}
\label{fig:samples}
\vspace{-0.4cm}
\end{figure}

\method-132K is a multilingual repo-level code completion dataset, covering four popular programming languages (\ie Python, Java, C++, and Go). It is collected from 2,000 high-quality open-source repositories and consists of 120,000 training and 12,000 testing samples. \method-132K has two features:
\begin{itemize}[leftmargin=*]
    \item \textbf{Long Cross-file Context.} Besides the in-file context, each sample is equipped with the long cross-file context. The average length of the cross-file context is about 12,000 tokens, and the maximum length is over 128,000 tokens. For comparison, a popular repo-level code completion benchmark - CrossCodeEval's context average 3390 tokens \cite{CrossCodeEval}.
    \item \textbf{Context-aware Code Completion.} To explicitly train LLMs to utilize the cross-file context, the code to be completed is highly context-aware. According to related works \cite{RepoCoder,GraphCoder,RLCoder}, we focus on two types of crucial information (\ie relevant APIs and similar code) in the cross-file context and ask LLMs to generate two types of context-aware code: (1) cross-file API invocations; (2) code spans similar to cross-file context. Figure~\ref{fig:samples} shows some code examples in \method-132K. Intuitively, to predict both types of code, LLMs need to master the capability of finding relevant information in the cross-file context, thus aligning LLMs with repo-level code completion.
\end{itemize}

\subsubsection{Data Collection Pipeline}
\label{sec:CoLT:data:collection}

The data collection pipeline of \method-132K consists of four stages:

\textbf{Stage 1: Code Repository Crawling.} In this stage, we crawl code repositories from GitHub. We design the following rules to ensure the quality and diversity of repositories:

\begin{itemize}[leftmargin=*]
    \item \textit{Multilingual Data}.
    Python, Java, C++, and Go are four popular programming languages in modern software development \cite{octoverse2024}. Thus, we prioritize repositories in these languages to establish a representative corpus.

    \item \textit{License Compliance}. 
    Repositories must adopt permissive open-source licenses (\eg MIT) to avoid legal issues. We use GHArchive \cite{GHArchive} to cross-check the license type and ensure compliance with permissive standards.

    \item \textit{Popularity}.
    We retain popular repositories that have at least 10 stars and updates within the last two years. Then we sort repositories based on the number of stars, commit frequency, and the presence of test files. The lowest 10\% of repositories are removed as they are considered low-quality.
\end{itemize}

To avoid data leakage, we exclude repositories in existing evaluation benchmarks (\ie CrossCodeEval in our experiments). For the test set of \method-132K, we only consider repositories created after March 2024, since the training data of the latest studied LLM is cut off at March 2024.

\textbf{Stage 2: Repo-level Deduplication.}  
Duplicate training data will lead to model overfitting \cite{DataDeduplication}. Thus, we conduct the repo-level data deduplication, which consists of two steps:

\begin{itemize}[leftmargin=*]
    \item \textit{Exact deduplication.} We remove repositories where all files have the same content.

    \item \textit{Near deduplication.} Exact deduplication may cause false negatives, so we further perform near deduplication. We compute the MinHash \cite{minhash} of all files with 256 permutations and use Locality Sensitive Hashing (LSH) \cite{lsh} to identify clusters of duplicates across repositories. We further reduce these clusters by ensuring that each file in the original cluster is similar to at least one other file in the reduced cluster. We consider two files near-duplicate when their Jaccard similarity \cite{jaccard1901etude} exceeds a threshold. We follow related work \cite{aixcoder} and empirically set the threshold to 0.85.
\end{itemize}

\textbf{Stage 3: Dependency Graph Construction.} We use SCIP \cite{scip} to parse these repositories. Repositories that fail to parse or exceed the time limit are excluded. The structured output of SCIP contains the locations of various code elements within the repository (\eg classes and functions) and the dependencies between these elements (\eg function invocations). This information can then support the subsequent stages.

\textbf{Stage 4: Context-aware Code Extraction.} In this stage, we extract the context-aware code as the ground-truth code (\ie code to be completed). As stated in Section \ref{sec:CoLT:data}, we focus on extracting two types of context-aware code:

\begin{itemize}[leftmargin=*]
    \item \textit{Cross-file API invocations.} Based on the dependency graphs built in stage 3, we identify code lines that contain cross-file API invocations. These lines are then extracted as the ground-truth code.

    \item \textit{Code spans similar to cross-file context.} Following previous work \cite{aixcoder}, we sample structured spans from repositories, ensuring they meet the following criteria: non-empty, not comments. We only include code spans, whose maximum similarity with retrieved code snippets in the cross-file context exceeds a threshold. How to retrieve code snippets is described in Stage 5. \changeline{We manually inspect some retrieved code candidates and empirically set the threshold to 0.3.}
\end{itemize}

To avoid data leakage, we exclude the ground-truth code appearing in our evaluation benchmarks (\ie CrossCodeEval in our experiments). 

\textbf{Stage 5: Context Extraction.} This stage is to extract the input context corresponding to the ground-truth code.
\begin{itemize}[leftmargin=*]
    \item \textit{In-file context.} Following previous work \cite{aixcoder}, we locate the current file (\ie the source file of the ground-truth code) and extract the prefix and suffix as in-file context. The prefix refers to the code preceding the ground-truth code, while the suffix is the code following the ground-truth code.

    \item \textit{Cross-file context:} Extracting cross-file contexts is an open problem in repo-level code completion. Following previous works \cite{CoCoMIC, RepoCoder, GraphCoder}, we adopt two mainstream strategies for extracting cross-file context:
    \begin{enumerate}[leftmargin=*]
        \item \textit{Dependency-based strategy.} The motivation is that imported files often contain elements used by the current file, such as functions. Thus, based on the dependency graphs built in Stage 3, we extract external files imported by the current file as cross-file context.

        \item \textit{Retrieval-based strategy.} The idea of this strategy is to retrieve similar code snippets across the repository as cross-file context. Following previous works \cite{RepoCoder,GraphCoder}, we divide the entire repository into multiple code snippets by splitting it into code snippets of fixed size. The size is empirically set to 20 lines according to previous works \cite{RepoCoder,GraphCoder}. These code snippets form a retrieval corpus. From the end of the in-file prefix, we extract a code snippet that serves as the query. We then calculate the Jaccard similarity between the query and each snippet in the retrieval corpus at the token level, ranking the snippets based on similarity. Following previous work \cite{RepoCoder}, we select the top 10 most similar code snippets as cross-file contexts.
    \end{enumerate}
\end{itemize}

\subsection{A Two-Stage Training Pipeline}
\label{sec:CoLT:RL}

\begin{algorithm}[t]
\caption{The two-stage training pipeline in \method.}
\label{alg:RL}
\KwIn{$\mathcal{M}$: An LLM to be trained, $\mathcal{D}_{\text{SFT}}$: Training data for SFT, $\mathcal{D}_{\text{RL}}$: Training data for RL}
\KwOut{$\mathcal{M}_{\text{CoLA}}$: Trained model using \method}
\BlankLine




\BlankLine
\underline{\textbf{Supervised Fine-Tuning (SFT):}}

Initialize model parameters $\mathcal{M}_{\text{SFT}}$ with $\mathcal{M}$\;
\For{each batch $B$ from $\mathcal{D}_{\text{SFT}}$}{
    Compute the loss $\mathcal{L}_{\text{SFT}}$ on the batch $B$ using Equation \ref{equ:loss_sft}\;
    Update $\mathcal{M}_{\text{SFT}}$ by minimizing $\mathcal{L}_{\text{SFT}}$\;
}
\BlankLine
\underline{\textbf{Preference Data Construction:}}

\textcolor{gray}{// \emph{Create preference data for RL training}}\;
$\mathcal{D}_{\text{Pre}} \leftarrow \emptyset$ \tcp*{Initialize preference dataset}
\For{each prompt $\mathbf{I}$ and ground truth $\mathbf{O}^*_i$ in $\mathcal{D}_{\text{DPO}}$}{
    $\mathbf{O} \leftarrow$ Generate multiple candidate completions using $\mathcal{M}_{\text{SFT}}(\mathbf{I}_i)$\;
    Filter $\mathbf{O}^-$ by removing empty, exact matches, partial matches, and duplicates\;
    \For{each $\mathbf{O}^-_j$ in $\mathbf{O}^-$}{
        $\mathcal{D}_{\text{Pre}} \leftarrow \mathcal{D}_{\text{Pre}} \cup \{(\mathbf{I}_i, \mathbf{O}^*_i, \mathbf{O}^-_j)\}$\;
    }
}
\BlankLine
\underline{\textbf{Reinforcement Learning (RL):}}

Initialize $\mathcal{M}_{\text{RL}}$ with $\mathcal{M}_{\text{SFT}}$\;
$\pi_{\text{ref}} \leftarrow \mathcal{M}_{\text{SFT}}$\;
\For{each batch $B$ in $\mathcal{D}_{\text{Pre}}$}{
    Compute the loss $\mathcal{L}_{\text{RL}}$ on the batch $B$ using Equation \ref{equ:loss_dpo}\;
    Update $\mathcal{M}_{\text{RL}}$ by minimizing $\mathcal{L}_{\text{RL}}$\;
}
$\mathcal{M}_{\text{CoLA}} \leftarrow \mathcal{M}_{\text{RL}}$;

\Return $\mathcal{M}_{\text{CoLA}}$
\end{algorithm}
\vspace{-0.1cm}

\method employs a two-stage training pipeline upon the dataset, including Supervised Fine-Tuning (SFT) and Reinforcement Learning (RL). We split the training data into two parts in a 5:5 ratio for two stages. An overview of the training pipeline is shown in Algorithm \ref{alg:RL}.

\subsubsection{Supervised Fine-Tuning (SFT, Line 1-6 in Algorithm \ref{alg:RL})} SFT is a common approach to aligning LLMs to specific tasks and is helpful to stabilize reinforcement learning. Through the SFT, models can know the repo-level code completion task and generate plausible completions. Specifically, we fine-tune models by minimizing the following loss function:
\begin{equation}
\label{equ:loss_sft}
    \mathcal{L}_{\text{SFT}} = -\frac{1}{N} \sum_{i=1}^{N} \log P(\mathbf{O}^*_i | \mathbf{I}_i),
\end{equation}
where $\mathbf{I}_i$ and $\mathbf{O}^*_i$ denote the input context and ground-truth code in the $i$-th training sample, respectively. $N$ is the size of the training data.

\subsubsection{Reinforcement Learning (RL, Line 7-21 in Algorithm \ref{alg:RL})} This stage aims to leverage RL to further improve the model's capability in utilizing cross-file context. 
We employ a popular offline RL approach - \textit{Direct Preference Optimization (DPO)} \cite{DPO}. DPO can refine the model's behavior without an explicit reward model and has shown impressive results in many tasks such as mathematical reasoning \cite{Step-DPO,Self-Play}.

Specifically, DPO first collects lots of preference data, which comprises triples - \{input context, chosen code, rejected code\}. The chosen code is the ground-truth code, and the rejected code is the incorrect code generated by models. DPO increases the generation probability of the chosen code and decreases the generation probability of the rejected code. To obtain the rejected code, we use the SFT-tuned model to generate multiple candidate completions for each input context. We select high-quality rejected code from these candidate completions through rigorous filtration. The filtration excludes three types of completions: (i) empty completions, (ii) same completions as ground truth code, and (iii) completions containing the ground-truth code.

Based on the preference data, we optimize the models by minimizing the following loss function:
\begin{equation}
\label{equ:loss_dpo}
    \mathcal{L}_{\text{RL}} = -\mathbb{E} \bigg[ \log \sigma\bigg(\beta \bigg(
\log \frac{\pi_\theta(\mathbf{O}^{*} | \mathbf{I})}{\pi_{\text{ref}}(\mathbf{O}^{*} | \mathbf{I})} 
- \log \frac{\pi_\theta(\mathbf{O}^{-} | \mathbf{I})}{\pi_{\text{ref}}(\mathbf{O}^{-} | \mathbf{I})}
\bigg)\bigg)\bigg]
\end{equation}
where $\mathbf{I}$, $\mathbf{O}^{*}$, and $\mathbf{O}^{-}$ denote the input context, chosen code, and rejected code. $\pi_\theta$ is the current model being optimized and $\pi_{\text{ref}}$ is the frozen reference model (\ie the SFT-tuned model). $\beta$ is a scaling hyper-parameter, and $\sigma$ is the common sigmoid activation function. 

The loss function encourages the model $\pi_\theta$ to assign higher probabilities to $\mathbf{O}^{*}$ and lower probabilities to $\mathbf{O}^{-}$. The term $\log \frac{\pi_\theta(\mathbf{O} | \mathbf{I}')}{\pi_{\text{ref}}(\mathbf{O} | \mathbf{I}')}$ represents the log-probability ratio between the current model and the reference model. This ratio acts as a regularization term, ensuring that the optimized model does not deviate too far from the original SFT-tuned policy.

\section{aiXcoder-7B-v2}
\label{sec:aiXcoder}

\method is general to existing LLMs. We take the SOTA LLM on code completion - aiXcoder-7B \cite{aixcoder} as an example and present  aiXcoder-7B-v2. This section describes the training details and the training curves. 

\subsection{Training Details}
\label{sec:aiXcoder:details}

\begin{figure*}[t]
  \centering
  \begin{subfigure}{0.32\linewidth}
    \includegraphics[width=1\textwidth]{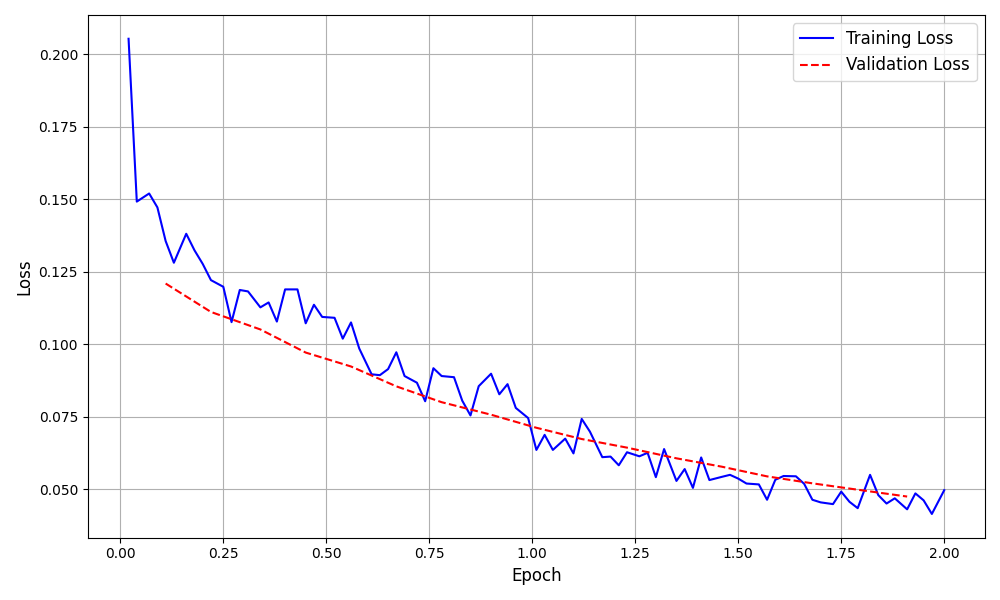}
    \caption{Training Losses During SFT}
    \label{fig:sft_loss}
  \end{subfigure}
  \hfill
  \begin{subfigure}{0.32\linewidth}
    \includegraphics[width=1\textwidth]{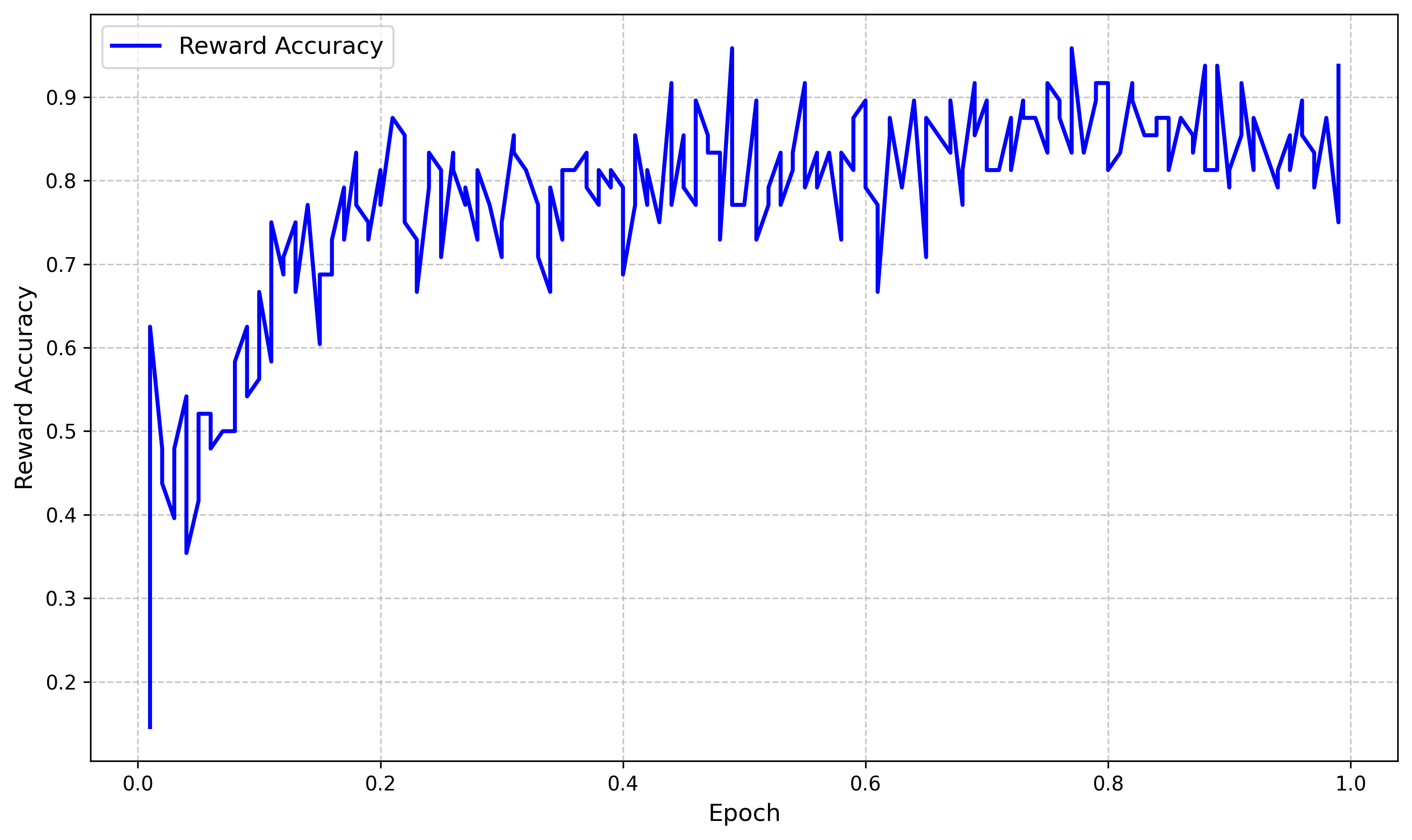}
    \caption{Reward Accuracy During DPO Training}
    \label{fig:dpo_reward_accuracy}
  \end{subfigure}
  \hfill
  \begin{subfigure}{0.32\linewidth}
    \includegraphics[width=1\textwidth]{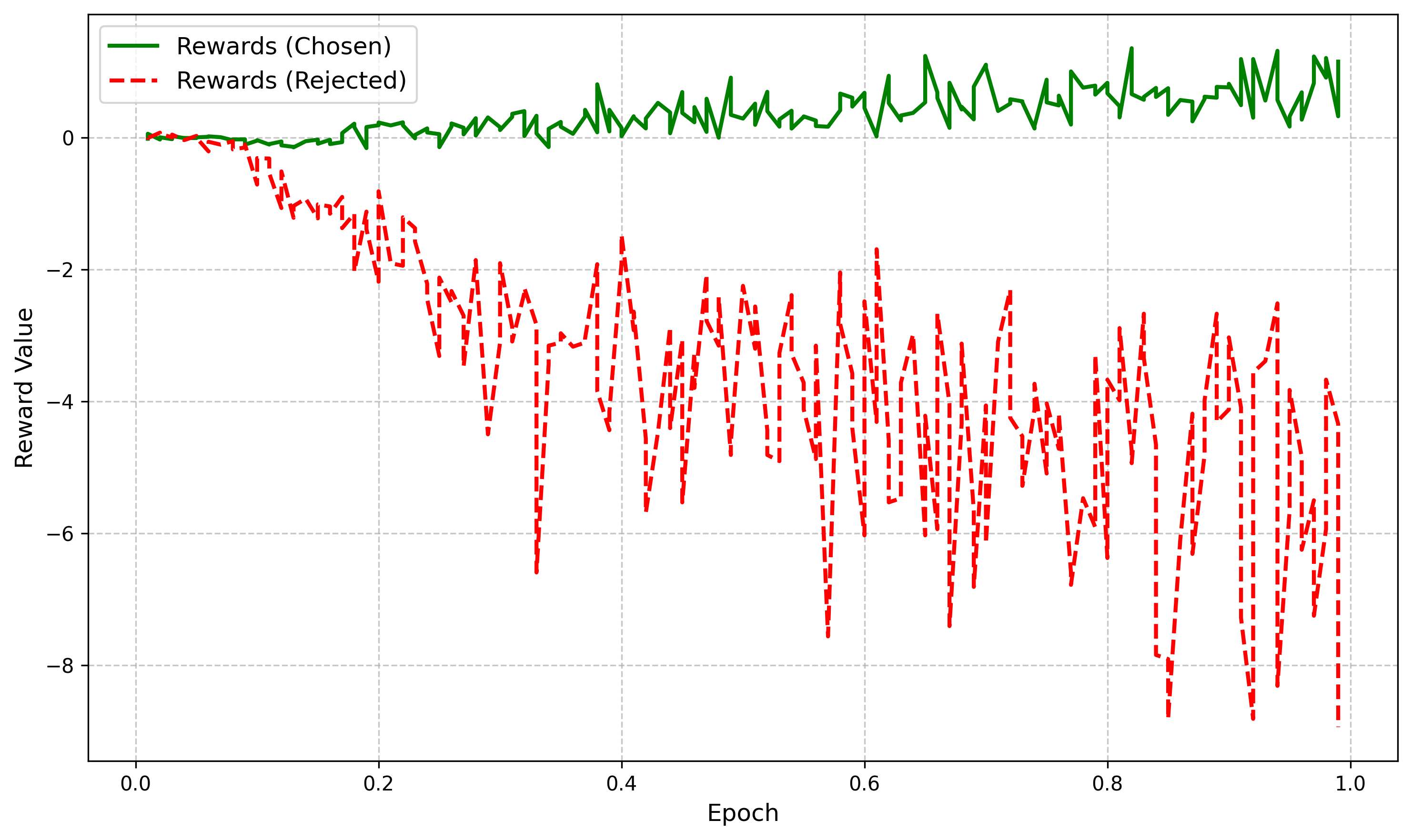}
    \caption{Rewards During DPO Training}
    \label{fig:dpo_chosen_rejected_rewards}
  \end{subfigure}
  \caption{Training curves in SFT and RL. (a) Training and validation losses during SFT. (b) Reward accuracy during RL training. (c) Chosen vs. rejected rewards during RL training.}
  \label{fig:training_curves}
  \vspace{-0.3cm}
\end{figure*}

\textbf{SFT (Line 1-6 in Algorithm \ref{alg:RL}).} We initialize $\mathcal{M}_{\text{SFT}}$ with aiXcoder-7B. Then, we fine-tune the model for 2 epochs. To prevent overfitting, we use a small learning rate of $2 \times 10^{-6}$, a weight decay of 0.01, and a maximum gradient norm of 1.0. The batch size is 128. These hyper-parameters are inspired by previous work \cite{Qwen2}. For computational efficiency, we employ the TRL framework \cite{trl} with DeepSpeed Zero-3 \cite{deepspeed} and Flash Attention-2 \cite{FlashAttention-2}. SFT is conducted on 8 NVIDIA A100 (40GB) GPUs and costs 20 hours.

\textbf{Preference Data Construction (Line 7-14 in Algorithm \ref{alg:RL}).} For each sample, the SFT-tuned model $\mathcal{M}_{\text{SFT}}$ generates 10 candidate completions using Top-p sampling \cite{Nucleus_Sampling} with $p=0.95$ and temperature $T=1.5$ to obtain diverse completions. To ensure training efficiency, we select up to three unique and incorrect completions for each sample as rejected code. Finally, we obtain 75,169 triples as preference data.

\textbf{DPO (Line 15-20 in Algorithm \ref{alg:RL}).} We implement the $\mathcal{M}_{\text{DPO}}$ and reference model $\pi_{\text{ref}}$ with the SFT-tuned model $\mathcal{M}_{\text{SFT}}$. To improve training efficiency, we use Low-Rank Adaptation (LoRA) \cite{LoRA} to train $\mathcal{M}_{\text{DPO}}$. The rank and alpha values of LoRA are set to 32 and 16. We apply LoRA to all transformer modules, including all self-attention weights and feed-forward weights. We train $\mathcal{M}_{\text{DPO}}$ for 1 epoch over the preference dataset with a batch size of 128. $\beta$ is set to 0.9. Existing work \cite{biderman2024lora} found that LoRA's best learning rates often in $[1 \times 10^{-5}, 5 \times 10^{-4}]$. Thus, we use a learning rate of $1 \times 10^{-4}$ with a linear scheduler and warm-up. DPO is conducted on 8 NVIDIA A100 (40GB) GPUs and costs 31 hours.

\subsection{Training Curves}
\label{sec:aiXcoder:curves}

We show the training curves of aiXcoder-7B-v2, including the training loss, reward accuracy, and the reward of chosen and rejected code.

\textbf{Training curves in SFT.} During the SFT phase, we randomly selected 5\% of the training data for validation. As shown in Figure~\ref{fig:sft_loss}, both the training and validation losses decrease steadily over the course of training, indicating that the model is converging.

\textbf{Training curves in RL.}
We monitor the RL training process using two metrics: reward accuracy and the reward of chosen and rejected code.
\ding{182} \textit{Reward accuracy} reflects the model's ability to distinguish between chosen and rejected code. The curve of reward accuracy is shown in Figure \ref{fig:dpo_reward_accuracy}. We can see a rapid increase in reward accuracy during the initial 0.4 epochs, where the accuracy rises from 0.2 to 0.8. Between 0.4 and 1.0 epochs, the accuracy continues to improve, reaching a plateau around 0.9. \ding{183} \textit{Reward of chosen and rejected code} means the mean log probabilities of the model for the chosen and rejected code. Figure \ref{fig:dpo_chosen_rejected_rewards} shows the reward of the chosen and rejected code during RL training. We can see that the reward of the chosen code consistently increases, and the reward of rejected completions declines. The two curves above show that the model effectively distinguishes between chosen and rejected code and tends to generate chosen code.

\section{Study Design}
\label{sec:study_design}

To evaluate the effectiveness of \method, we conduct a large-scale study. In this section, we describe the design of our study, including research questions, benchmarks, baselines, and evaluation metrics.

\subsection{Research Questions}
\label{sec:study_design:RQs}

Our study aims to answer the following research questions (RQs):

\textbf{RQ1: How much improvement does \method achieve in repo-level code completion?}
We aim to evaluate the improvement in repo-level code completion performance when using \method. We apply \method to aiXcoder-7B and present aiXcoder-7B-v2. Then we compare it to existing SOTA LLMs on multiple benchmarks.

\textbf{RQ2: How effectively does \method generalize across different programming languages?} This RQ investigates whether the context utilization capabilities learned by \method can generalize across languages. We train aiXcoder-7B-v2 on \method-132K covering four languages and test the model on the other two languages.

\textbf{RQ3: Does \method maintain its effectiveness across different LLMs?} 
Besides the aiXcoder-7B, we apply \method to other two popular LLMs (\ie DeepSeek-Coder-6.7B and Code Llama-7B) to evaluate whether \method can improve them in repo-level code completion.

\textbf{RQ4: Does \method improve the capability of LLMs to utilize the cross-file context?}
This RQ evaluates whether \method enhances the LLMs' capability of utilizing APIs and similar code in the cross-file context.

\textbf{RQ5: How do different types of context-aware code and training stages contribute to \method ?} This RQ conducts an ablation study, which explores the contributions of two types of context-aware code (\ie cross-file API invocations and code spans similar to cross-file context) and two training stages (\ie SFT and RL) to \method.

\subsection{Benchmarks}
\label{sec:study_design:benchmarks}

We conduct experiments on the following benchmarks:

\textbf{CoLT-132K (Test)} is the test set of \method-132K, consisting of 12,000 samples. It covers four programming languages (\ie Python, Java, C/C++, Go) and asks LLMs to generate two types of context-aware code, including cross-file API invocations and code spans similar to the cross-file context. Each language contains 3,000 samples. Notably, the test set was meticulously collected from repositories created after March 2024 to prevent data leakage. For detailed information, please refer to Section~\ref{sec:CoLT:data}.

\textbf{CrossCodeEval} \cite{CrossCodeEval} is a multilingual repo-level code completion benchmark, including 2,665 Python samples, 2,139 Java samples, 3,356 TypeScript samples, and 1,768 C\# samples. We reuse the cross-file context retrieved by BM25 provided in CrossCodeEval.

\subsection{Baseline Models}
\label{sec:study_design:baseline}

We selected 8 popular open-source LLMs for comparison, which are widely used in code completion tasks. These models are categorized into two groups based on their scales.

\ding{182} \textbf{LLMs with 7 billion parameters.} These LLMs have a comparable parameter scale to aiXcoder-7B-v2.

\begin{itemize}[leftmargin=*]
    \item \textbf{aiXcoder-7B} \cite{aixcoder} is developed by aiXcoder company. With 7 billion parameters, aiXcoder-7B achieves SOTA performance in coding tasks, particularly in code completion. 

    \item \textbf{aiXcoder-7B-cont} is a continually pre-trained model of aiXcoder-7B. We randomly sample code files from repositories in \method-132K and train aiXcoder-7B with its original pre-training objective. The amount of training tokens is equal to that of aiXcoder-7B-v2. We contacted the developers of aiXcoder-7B and used the official code implementation they provided for pre-training.
    
    \item \textbf{Qwen2.5-Coder-7B} \cite{Qwen2.5-Coder}, developed by Alibaba, is an enhanced version of Qwen2.5 \cite{Qwen2}. It is trained on an expansive dataset comprising source code, text-code alignment data, and synthetic data, amounting to 5.5 trillion tokens.

    \item \textbf{DeepSeek-Coder-6.7B} \cite{DeepSeek-Coder}, from DeepSeek AI, is trained from scratch on 2 trillion tokens across 87 programming languages. It is designed to handle complex coding tasks.
    
    \item \textbf{Code Llama-7B} \cite{CodeLlama}, developed by Meta AI, is initialized with Llama 2 \cite{llama2} and trained on 500B tokens from a large-scale code corpus. It supports common programming languages and has a default context length of 100K tokens.
    
    \item \textbf{StarCoder2-7B} \cite{Starcoder2}, introduced by BigCode, is trained on The Stack v2 dataset following additional code fine-tuning stage. It supports multilingual code generation.
    
\end{itemize}

\ding{183} \textbf{LLMs with larger sizes.} We also add two LLMs with larger scales as baselines.

\begin{itemize}[leftmargin=*]

    \item \textbf{Code Llama-13B} \cite{CodeLlama}, the upgraded version of the Code Llama model, expands the parameter size to 13 billion. 
    
    \item \textbf{DeepSeek-Coder-33B} \cite{DeepSeek-Coder} is the largest variant of the dense model of the DeepSeek-Coder series.
    
\end{itemize}

\begin{change}
We employ the official prompting formats of LLMs for fill-in-the-middle (FIM) code completion. Specifically,
\begin{itemize}[leftmargin=*]
    \item DeepSeek-Coder:\texttt{[cross-file context] <|fim\_begin|>[prefix]<|fim\_hole|> [suffix]<|fim\_end|>}
    
    \item aiXcoder:\texttt{[cross-file context]<AIX-PRE> <AIX-POST>[suffix]\allowbreak<AIX-MIDDLE>[prefix]}
    
    \item Code Llama:\texttt{[cross-file context]<PRE>[pre- fix]<SUF>[suffix]<MID>}
    
    \item Qwen2.5-Coder:\texttt{<|repo\_name|>[name]<|file\_ sep|>[path]<|fim\_prefix|>[prefix] <|fim\_suffix|>[suffix]<|fim\_middle|>}
\end{itemize}

When producing prompts, we first include the in-file prefix and suffix. Then, we add cross-file context from dependency-based and retrieval-based methods in an alternating order, until reaching the maximum context window of LLMs. We employ the original maximum context window of each model.

We omit some related repo-level code completion approaches for comparison\cite{RepoCoder,GraphCoder,CoCoMIC,RLCoder}, as we focus on different research problems. \textbf{Our approach focuses on improving the capability of LLMs in utilizing context.} In contrast, existing repo-level code completion approaches \cite{RepoCoder,GraphCoder,CoCoMIC,RLCoder} focus on how to extract relevant context and treat LLMs as black-box code generators. In practice, our work and these recent works are complementary. The experimental results in Section \ref{sec:discussion:complement} demonstrate the complementarity of our work and existing repo-level code completion approaches.

\end{change}

\subsection{Evaluation Metrics}
\label{sec:study_design:metrics}

Following previous works \cite{RepoCoder,CrossCodeEval,CoCoMIC,aixcoder}, we use the following two evaluation metrics:

\begin{itemize}[leftmargin=*]
    \item \textbf{Exact Match (EM)} quantifies the proportion of predictions that identically match ground truths.
    \item \textbf{BLEU \cite{Bleu}} computes $n$-gram overlap between predictions and ground truths. $n$ is empirically set to 4.
\end{itemize}

Following recent works \cite{DeepSeek-Coder, CrossCodeEval}, we use greedy decoding during inference. When predicting API invocations, we extract the first non-empty line of LLMs' completions for evaluation. When predicting code spans, we truncate the model outputs at the special token and extract the rest code for evaluation. 

\begin{table*}[t]
    \centering
    \caption{The performance of LLMs in \method-132K.}
    \setlength{\tabcolsep}{2pt}
    \label{tab:rq1_api_line}
    \resizebox{\textwidth}{!}{
    \begin{tabular}{lcccccccccccccccc}
        \toprule
        \multirow{3}{*}{\textbf{Model}} & 
        \multicolumn{8}{c}{\textbf{Cross-file API Invocation}} & 
        \multicolumn{8}{c}{\textbf{Code Span}} \\
        \cmidrule(lr){2-9} \cmidrule(lr){10-17}
        & \multicolumn{2}{c}{Python} & \multicolumn{2}{c}{Java} & \multicolumn{2}{c}{C++} & \multicolumn{2}{c}{Go} &
        \multicolumn{2}{c}{Python} & \multicolumn{2}{c}{Java} & \multicolumn{2}{c}{C++} & \multicolumn{2}{c}{Go} \\
        \cmidrule(lr){2-3} \cmidrule(lr){4-5} \cmidrule(lr){6-7} \cmidrule(lr){8-9}
        \cmidrule(lr){10-11} \cmidrule(lr){12-13} \cmidrule(lr){14-15} \cmidrule(lr){16-17}
        & EM & BLEU & EM & BLEU & EM & BLEU & EM & BLEU & EM & BLEU & EM & BLEU & EM & BLEU & EM & BLEU \\
        
        \midrule
        \multicolumn{17}{c}{\textbf{7B Models}} \\
        \midrule
        
        Code Llama-7B 
        & 55.8 & 68.4 & 66.7 & 78.5 & 50.7 & 65.6 & 50.1 & 63.4 
        & 39.9 & 56.8 & 53.9 & 68.5 & 37.3 & 53.1 & 46.5 & 61.2 \\
        
        StarCoder2-7B 
        & 46.9 & 61.3 & 64.8 & 78.4 & 37.8 & 52.9 & 51.2 & 69.6 
        & 25.5 & 37.5 & 40.3 & 50.2 & 24.8 & 36.8 & 34.8 & 47.2 \\
        
        DeepSeek-Coder-6.7B 
        & 58.0 & 69.4 & 67.4 & 78.6 & 46.8 & 59.7 & 57.5 & 71.7 
        & 41.4 & 59.1 & 58.9 & 73.2 & 38.5 & 54.8 & 57.2 & 72.2 \\
        
        Qwen2.5-Coder-7B 
        & 55.3 & 75.6 & 66.1 & 84.4 & 50.7 & 74.0 & 45.8 & 69.5 
        & 48.1 & 67.9 & 59.1 & 81.6 & 46.5 & 68.7 & 57.3 & 75.9 \\
        
        aiXcoder-7B 
        & 62.5 & 78.5 & 73.3 & 88.7 & 56.0 & 79.2 & 62.3 & 83.2 
        & 48.6 & 67.1 & 68.8 & 83.4 & 52.9 & 70.1 & 65.2 & 79.0 \\

        aiXcoder-7B-cont
        & 61.6 & 73.7 & 73.5 & 84.6 & 56.0 & 71.3 & 61.7 & 76.8
        & 42.9 & 59.2 & 67.1 & 80.6 & 47.2 & 62.8 & 47.1 & 59.7 \\
        
        \rowcolor[rgb]{ .741,  .843,  .933} aiXcoder-7B-v2 
        & \textbf{69.7} & \textbf{83.8} & \textbf{78.5} & \textbf{89.6} & \textbf{64.1} & \textbf{80.6} & \textbf{68.2} & \textbf{84.1} 
        & \textbf{58.2} & \textbf{76.7} & \textbf{74.8} & \textbf{89.2} & \textbf{61.1} & \textbf{79.9} & \textbf{73.8} & \textbf{87.4} \\
        
        \midrule

        \multicolumn{1}{c}{\textbf{Relative Improve.}} & +11.5\% & +6.8\% & +7.1\% & +1.0\% & +14.5\% & +1.8\% & +9.5\% & +1.1\% 
        & +19.7\% & +14.3\% & +8.8\% & +7.0\% & +15.5\% & +13.9\% & +13.2\% & +10.6\% \\
        
        \midrule
        \multicolumn{17}{c}{\textbf{$>$7B Models}} \\
        \midrule
        
        Code Llama-13B 
        & 56.4 & 69.8 & 67.2 & 79.2 & 52.1 & 67.5 & 45.4 & 55.9 
        & 46.1 & 63.1 & 61.6 & 76.6 & 45.8 & 62.7 & 49.6 & 66.5 \\
        
        DeepSeek-Coder-33B 
        & 61.3 & 71.6 & 70.1 & 81.2 & 51.5 & 65.6 & 60.1 & 73.9 
        & 44.6 & 62.5 & 62.4 & 77.2 & 43.8 & 61.0 & 59.4 & 74.8 \\
        
        \bottomrule
    \end{tabular}}
\end{table*}

\section{Results and Analyses}
\label{sec:results}

\subsection{RQ1: Improvements in Repo-level Code Completion}
\label{sec:results:rq1}

\textbf{Setting.} As stated in Section \ref{sec:aiXcoder}, we apply \method to aiXcoder-7B and present aiXcoder-7B-v2. To validate the effectiveness of \method, we compare aiXcoder-7B-v2 to eight LLMs on repo-level code completion.

\textbf{Results and Analyses.} The results are shown in Table \ref{tab:rq1_api_line} and Table \ref{tab:rq1_cceval_python_java}. We show the relative improvements of aiXcoder-7B-v2 compared to aiXcoder-7B.

\textbf{\method substantially improves the performance of aiXcoder-7B on repo-level code completion.} aiXcoder-7B-v2 substantially outperforms aiXcoder-7B on \method-132K and CrossCodeEval. Especially when completing Python’s code spans, aiXcoder-7B-v2 outperforms aiXcoder-7B by 19.7\% in EM and 14.3\% in BLEU. aiXcoder-7B-v2 even outperforms larger models, including Code Llama-13B and DeepSeek-Coder-33B. The results show that \method can effectively improve the performance of LLMs in repo-level code completion.

\textbf{The effectiveness of \method is not attributed to additional training data.} aiXcoder-7B-cont is continually trained from aiXcoder-7B with its original pre-training objectives. The amount of training tokens is equal to that of aiXcoder-7B-v2. We can see that aiXcoder-7B-v2 substantially outperforms aiXcoder-7B-cont. It demonstrates that the superiority of \method is not caused by additional training data. We also notice that aiXcoder-7B-cont is slightly weaker than aiXcoder-7B. We speculate that this is because aiXcoder-7B has been fully trained on 1.2 trillion unique tokens. Further pre-training may cause slight overfitting. We also contacted the development team of aiXcoder-7B and confirmed the reliability of the experimental results.

\begin{tcolorbox}[size=title]
    \textbf{Answer to RQ1:} aiXcoder-7B-v2 substantially improves aiXcoder-7B on repo-level code completion, showing \method's effectiveness. The improvements over aiXcoder-7B-cont exclude the impact of additional training data.
\end{tcolorbox}

\subsection{RQ2: Generalization Across Different Programming Languages}
\label{sec:results:rq2}

\textbf{Setting.} We train aiXcoder-7B-v2 on \method-132K including four programming languages (\ie Python, Java, C++, and Go). To evaluate generalization ability, we test it on the other two languages (\ie C\# and TypeScript) in CrossCodeEval.

\begin{table}[t]
    \centering
    \caption{The performance of LLMs on CrossCodeEval.}
    \label{tab:rq1_cceval_python_java}
    \resizebox{0.9\linewidth}{!}{
    \begin{tabular}{lcccc}
        \toprule
        \multirow{2}{*}{\textbf{Model}} & 
        \multicolumn{2}{c}{\textbf{Python}} & 
        \multicolumn{2}{c}{\textbf{Java}} \\
        \cmidrule(lr){2-3} \cmidrule(lr){4-5}
        & EM & BLEU & EM & BLEU \\
        
        \midrule
        \multicolumn{5}{c}{\textbf{7B Models}} \\
        \midrule

        Code Llama-7B 
        & 36.06 & 54.27 & 40.58 & 65.04 \\

        StarCoder2-7B 
        & 11.97 & 37.04 & 34.32 & 56.71 \\

        DeepSeek-Coder-6.7B 
        & 13.32 & 35.00 & 36.65 & 59.38 \\
        
        Qwen2.5-Coder-7B 
        & 38.27 & 57.53 & 42.59 & 65.98 \\
        
        aiXcoder-7B 
        & 31.22 & 52.69 & 42.50 & 66.59 \\

        aiXcoder-7B-cont
        & 32.26 & 52.86 & 43.19 & 66.72 \\
        
        \rowcolor[rgb]{ .741,  .843,  .933} aiXcoder-7B-v2 
        & \textbf{40.52} & \textbf{63.40} & \textbf{51.43} & \textbf{74.95} \\
        
        \midrule
        \multicolumn{1}{c}{\textbf{Relative Improve.}}
        & +29.8\% & +20.3\% & +21.0\% & +12.6\% \\
        
        \midrule
        \multicolumn{5}{c}{\textbf{$>$7B Models}} \\
        \midrule
        
        Code Llama-13B 
        & 38.76 & 57.19 & 42.87 & 66.01 \\
        
        DeepSeek-Coder-33B 
        & 18.46 & 39.39 & 37.21 & 59.57 \\
        
        \bottomrule
    \end{tabular}}
\end{table}

\begin{table}[t]
    \centering
    \caption{The generalization ability of aiXcoder-7B-v2.}
    \label{tab:rq2_aixcoder}
    \resizebox{0.9\linewidth}{!}{
    \begin{tabular}{lcccc}
        \toprule
        \multirow{2}{*}{\textbf{Model}} & 
        \multicolumn{2}{c}{\textbf{TypeScript}} & 
        \multicolumn{2}{c}{\textbf{C\#}} \\
        \cmidrule(lr){2-3} \cmidrule(lr){4-5}
        & EM & BLEU & EM & BLEU \\
        
        \midrule
        \multicolumn{5}{c}{\textbf{7B Models}} \\
        \midrule
        
        Code Llama-7B & 45.29 & 61.53 & 51.07 & 68.22 \\
        StarCoder2-7B & \changeline{33.55} & \changeline{55.65} & \changeline{42.48} & \changeline{61.47} \\
        DeepSeek-Coder-6.7B & 34.80 & 53.02 & 37.05 & 52.44 \\
        Qwen2.5-Coder-7B & 47.34 & 66.28 & \textbf{57.86} & 74.28 \\
        aiXcoder-7B
        & 46.96 & 66.65 & 46.32 & 68.97 \\

        aiXcoder-7B-cont
        & 47.19 & 67.24 & 47.16 & 69.62 \\
        
        \rowcolor[rgb]{ .741,  .843,  .933} aiXcoder-7B-v2
        & \textbf{55.96} & \textbf{75.74} & 55.88 & \textbf{76.17} \\

        \midrule
        
        \multicolumn{1}{c}{\textbf{Relative Improve.}}
        & +19.2\% & +13.6\% & +20.6\% & +10.4\% \\
                
        \midrule
        \multicolumn{5}{c}{\textbf{$>$7B Models}} \\
        \midrule
        
        Code Llama-13B & 49.20 & 64.55 & 57.07 & 71.22 \\
        DeepSeek-Coder-33B & 38.17 & 56.32 & 38.12 & 53.95 \\

        \bottomrule
    \end{tabular}}
\end{table}

\textbf{Results and Analyses.} The results are shown in Table \ref{tab:rq2_aixcoder}.

\textbf{The context utilization ability learned by \method can generalize to new languages.} In languages not included in \method-132K, such as C\# and TypeScript, aiXcoder-7B-v2 demonstrates strong performance. In TypeScript, EM increases from 46.96 to 55.96, a 19.17\% improvement. In C\#, EM increases from 46.32 to 55.88, a boost of 20.64\%. These results show that the context utilization capability learned by \method has a good generalization ability.

\begin{tcolorbox}[size=title]
    \textbf{Answer to RQ2:} The capability learned by \method generalizes to new languages. In languages like C\# and TypeScript, aiXcoder-7B-v2 achieves significant improvements, with EM increasing by up to 19.17\% and 20.64\%.
\end{tcolorbox}

\subsection{RQ3: Effectiveness Across Different LLMs}
\label{sec:results:rq3}
\begin{table*}[t]
    \centering
    \setlength{\tabcolsep}{2pt}
    \caption{The performance of \method on DS-Coder-6.7B and Code Llama-7B upon \method-132K. ``DS'': DeepSeek.}
    \label{tab:rq3}
    \resizebox{\textwidth}{!}{
    \begin{tabular}{lcccccccccccccccc}
        \toprule
        \multirow{3}{*}{\textbf{Model}} & 
        \multicolumn{8}{c}{\textbf{Cross-file API Invocation}} & 
        \multicolumn{8}{c}{\textbf{Code Span}} \\
        \cmidrule(lr){2-9} \cmidrule(lr){10-17}
        & \multicolumn{2}{c}{Python} & \multicolumn{2}{c}{Java} & \multicolumn{2}{c}{C++} & \multicolumn{2}{c}{Go} &
        \multicolumn{2}{c}{Python} & \multicolumn{2}{c}{Java} & \multicolumn{2}{c}{C++} & \multicolumn{2}{c}{Go} \\
        \cmidrule(lr){2-3} \cmidrule(lr){4-5} \cmidrule(lr){6-7} \cmidrule(lr){8-9}
        \cmidrule(lr){10-11} \cmidrule(lr){12-13} \cmidrule(lr){14-15} \cmidrule(lr){16-17}
        & EM & BLEU & EM & BLEU & EM & BLEU & EM & BLEU & EM & BLEU & EM & BLEU & EM & BLEU & EM & BLEU \\

        \midrule

        DS-Coder-6.7B 
        & 58.0 & 69.4 & 67.4 & 78.6 & 46.8 & 59.7 & 57.5 & 71.7 
        & 41.4 & 59.1 & 58.9 & 73.2 & 38.5 & 54.8 & 57.2 & 72.2 \\
        
        \rowcolor[rgb]{ .741,  .843,  .933} \textbf{\quad +\method} 
        & \textbf{70.7} & \textbf{84.3} & \textbf{77.0} & \textbf{88.7} & \textbf{63.3} & \textbf{78.9} & \textbf{68.2} & \textbf{84.1} 
        & \textbf{57.6} & \textbf{75.6} & \textbf{73.6} & \textbf{87.6} & \textbf{59.2} & \textbf{77.8} & \textbf{73.4} & \textbf{87.0} \\
        
        \multicolumn{1}{c}{\textbf{Relative Improve.}} 
        & +21.9\% & +21.5\% & +14.2\% & +12.9\% & +35.3\% & +32.1\% & +18.7\% & +17.3\% 
        & +39.1\% & +27.9\% & +25.0\% & +19.7\% & +53.8\% & +41.9\% & +28.4\% & +20.4\% \\
        
        \midrule
        
        Code Llama-7B 
        & 55.8 & 68.4 & 66.7 & 78.5 & 50.7 & 65.6 & 50.1 & 63.4 
        & 39.9 & 56.8 & 53.8 & 68.5 & 37.3 & 53.0 & 46.5 & 61.2 \\
        
        \rowcolor[rgb]{ .741,  .843,  .933} \textbf{\quad +\method} 
        & \textbf{67.8} & \textbf{84.5} & \textbf{75.7} & \textbf{87.1} & \textbf{60.3} & \textbf{77.2} & \textbf{66.1} & \textbf{82.7} 
        & \textbf{53.2} & \textbf{72.4} & \textbf{72.2} & \textbf{87.3} & \textbf{57.1} & \textbf{76.2} & \textbf{71.6} & \textbf{85.8} \\
        
        \multicolumn{1}{c}{\textbf{Relative Improve.}} 
        & +21.5\% & +23.5\% & +13.5\% & +10.9\% & +18.9\% & +17.7\% & +31.9\% & +30.4\% 
        & +33.3\% & +27.5\% & +34.2\% & +27.4\% & +53.2\% & +43.8\% & +54.0\% & +40.2\% \\
        
        \bottomrule
    \end{tabular}}
\end{table*}

This RQ validates that \method is general to different LLMs and can bring stable improvements.

\textbf{Setting.} We apply \method to DeepSeek-Coder-6.7B and Code Llama-7B. The training details are consistent with aiXcoder-7B-v2 (Section~\ref{sec:aiXcoder}). Then, we evaluate the improvements of \method upon both LLMs.

\textbf{Results and Analyses.} Table~\ref{tab:rq3} shows the results on \method-132K.

\textbf{\method brings substantial improvements on both LLMs.}
The application of \method to both LLMs brings significant improvements in both EM and BLEU. For DeepSeek-Coder-6.7B, the EM is improved by up to 53.8\% and BLEU is improved by up to 41.9\%. For Code Llama-7B, the EM is improved by up to 54\% and BLEU is improved by up to 43.8\%. These results show the effectiveness of \method across different LLMs. 

\begin{tcolorbox}[size=title]
    \textbf{Answer to RQ3:} Besides aiXcoder-7B, \method effectively improves different LLMs (\eg DeepSeek-Coder-6.7B and Code Llama-7B). The EM is improved by up to 54\% and BLEU is improved by up to 43.8\%.
\end{tcolorbox}


\subsection{RQ4: Improvements on Context Utilization Capability}
\label{sec:results:rq4}

In this RQ, we design two probing experiments to evaluate whether \method enhances the model’s ability to utilize APIs and similar code in the cross-file context.

\textbf{\ding{182} The performance in utilizing APIs in the cross-file context.}
\label{sec:results:rq4_api}

\begin{table}[t]
    \centering
    \caption{The API accuracy of LLMs on \method-132K.}
    \label{tab:rq4_api}
    \vspace{-0.2cm}
    \resizebox{0.8\linewidth}{!}{
    \begin{tabular}{lcccc}
        \toprule
        \multirow{2}{*}{\textbf{Model}} & \multicolumn{4}{c}{\textbf{API Accuracy (\%)}} \\
        \cmidrule(lr){2-5}
        & Python & Java & C++ & Go \\
        \midrule
        Code Llama-7B & 76.6 & 79.6 & 70.0 & 62.9 \\
        StarCoder2-7B & 71.0 & 81.3 & 57.6 & 69.1 \\
        DeepSeek-Coder-6.7B & 77.6 & 80.4 & 64.3 & 70.3 \\
        Qwen2.5-Coder-7B & 82.8 & 80.7 & 70.2 & 60.6 \\
        Code Llama-13B & 78.6 & 81.6 & 71.8 & 56.5 \\
        DeepSeek-Coder-33B & 79.7 & 82.8 & 69.8 & 74.2 \\
        aiXcoder-7B & 83.4 & 86.5 & 76.6 & 76.4 \\
        
        \rowcolor[rgb]{ .741,  .843,  .933} aiXcoder-7B-v2 & \textbf{94.4} & \textbf{92.4} & \textbf{87.7} & \textbf{81.3} \\
        \bottomrule
    \end{tabular}}
    \vspace{-0.2cm}
\end{table}

\textbf{Setting.} We expect LLMs can actively locate and invoke relevant APIs presented in the cross-file context. To quantify the model’s ability to invoke these APIs, we introduce a new metric: \textbf{API Accuracy (API Acc)}. It only measures the accuracy of API names in the completion results. If the API name in LLMs' completions matches the ground truth, it is counted as 1; otherwise, it is 0. Different from traditional evaluation metrics (\eg EM), API Acc only cares whether LLMs invoke correct APIs and excludes the impact of other factors (\eg arguments).

\textbf{Results and Analyses.} As shown in Table~\ref{tab:rq4_api}, across all four programming languages, \method substantially improves the API acc of aiXcoder-7B. Specifically, aiXcoder-7B-v2 surpasses aiXcoder-7B by 13.2\% on Python, 6.8\% on Java, 14.5\% on C++, and 6.4\% on Go. Figure \ref{fig:motivating_ex1} shows a real case, where only aiXcoder-7B-v2 correctly invokes APIs among the three LLMs. These results demonstrate that \method improves the capability of LLMs in utilizing relevant APIs.

\textbf{\ding{183} The performance in utilizing similar code in the cross-file context.}

\textbf{Setting.} In this experiment, we explore whether \method enhances the model's ability to utilize similar code. We categorize test samples in \method-132K (Code Spans) into different groups based on the maximum similarity between the \textbf{retrieved code snippets} and the \textbf{ground-truth code}. For each group, we report the number of correct completions by aiXcoder-7B and aiXcoder-7B-v2. 

\begin{figure}[htbp]
\centering
\includegraphics[width=\linewidth]{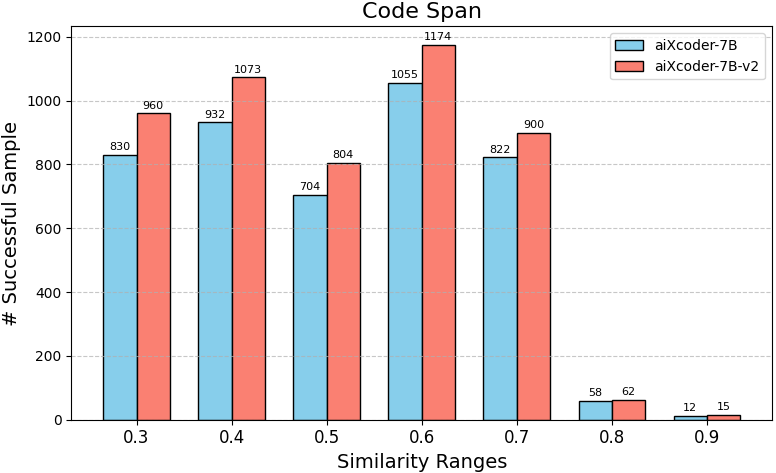}
\vspace{-0.2cm}
\caption{The performance of aiXcoder-7B and aiXcoder-7B-v2 on different groups of test samples. The x-axis represents the maximum similarity between the ground truth code and the code snippets.}
\label{fig:all_similar}
\vspace{-0.3cm}
\end{figure}

\textbf{Results and Analyses.} The results are presented in Figure~\ref{fig:all_similar}. aiXcoder-7B-v2 substantially outperforms aiXcoder-7B across all groups. Figure \ref{fig:motivating_ex2} shows a real case, where aiXcoder-7B-v2 successfully utilizes the information in similar code and generates correct code. The results show that \method enhances the ability of LLMs to reuse knowledge (\eg user-defined paths) in similar code.

\begin{tcolorbox}[size=title]
    \textbf{Answer to RQ4:} \method significantly improves the capability of aiXcoder-7B in utilizing relevant APIs and similar code in the cross-file context.
\end{tcolorbox}

\subsection{RQ5: Ablation study}
\label{sec:results:rq5}

In this RQ, we conduct an ablation study, which evaluates the contributions of two types of training data (\ie cross-file API invocations and code spans similar to cross-file context) and two training stages (\ie SFT and RL) to \method.

\textbf{Settings.} To quantify the contributions of each component, we remove different components from aiXcoder-7B-v2 and evaluate the model on \method-132K. We report the average EM and BLEU scores on the whole test set. Since DPO relies on SFT for cold start, otherwise it cannot stabilize training, we do not show the experimental results of utilizing only DPO.

\textbf{Results and Analyses.} The results of the ablation study are shown in Table \ref{tab:ablation_avg}. From the perspective of training stages, both SFT and DPO contribute to \method's performance. From the view of the training data, training only on cross-file API invocations or code spans causes slight overfitting, resulting in decreased accuracy. Thus, two types of context-aware code and two training stages all contribute to \method's performance.


\begin{table}[t]
\centering
\caption{The results of the ablation study.}
\vspace{-0.2cm}
\label{tab:ablation_avg}
\resizebox{0.8\columnwidth}{!}{
\begin{tabular}{ccccc}
\toprule
\textbf{SFT} & \textbf{DPO} & \textbf{EM} & \textbf{BLEU} \\
\midrule

\ding{55} & \ding{55} & 60.4 & 77.4 \\
\ding{51} & \ding{55} & 67.3 ($\uparrow$ 11.39\%) & 83.3 ($\uparrow$ 7.66\%) \\

\rowcolor[rgb]{ .741,  .843,  .933} \ding{51} & \ding{51} & \textbf{68.0} ($\uparrow$ 12.55\%) & \textbf{83.7} ($\uparrow$ 8.14\%) \\

\midrule
\textbf{API} & \textbf{Code Span} & \textbf{EM} & \textbf{BLEU} \\
\midrule

\ding{55} & \ding{55} & 60.4 & 77.4 \\
\ding{51} & \ding{55} & 52.4 ($\downarrow$ 13.2\%) & 72.1 ($\downarrow$ 6.8\%) \\

\ding{55} & \ding{51} & 60.2 ($\downarrow$ 0.3\%) & 76.7 ($\downarrow$ 0.9\%) \\

\rowcolor[rgb]{ .741,  .843,  .933} \ding{51} & \ding{51} & \textbf{68.0} ($\uparrow$ 12.55\%) & \textbf{83.7} ($\uparrow$ 8.14\%) \\

\bottomrule
\end{tabular}}
\vspace{-0.4cm}
\end{table}

\begin{tcolorbox}[size=title]
    \textbf{Answer to RQ5:} Two types of context-aware code and two training stages all contribute to the performance of \method.
\end{tcolorbox}

\begin{change}
\section{Discussion}

\subsection{Performance on Single-file Code Completion and Code Generation}
\label{sec:discussion:othertasks}

\vspace{-0.3cm}
\begin{table}[ht]
    \centering
    \changeref 
    \caption{\changeline{The performance of aiXcoder-7B and aiXcoder-7B-v2 upon single-file completion (EM and BLEU) and code generation (Pass@1).}}
    \label{D1:single_humaneval}
    \begin{tabular}{lccc}
        \toprule
        \multirow{2}{*}{\textbf{Model}} & \multicolumn{2}{c}{\textbf{Single-file}} & \textbf{HumanEval} \\
        \cmidrule(lr){2-3} \cmidrule(lr){4-4}
        & EM & BLEU & Pass@1 \\
        \midrule
        aiXcoder-7B & 46.1 & 65.8 & 43.2 \\
        \rowcolor[rgb]{ .741,  .843,  .933} aiXcoder-7B-v2 & 53.1 & 74.6 & \textbf{51.8} \\
        \textbf{Relative Improve.} & +15.2\% & +13.8\% & +20.0\% \\
        \bottomrule
    \end{tabular}
    \vspace{-0.3cm}
\end{table}

\begin{table}[t]
    \centering
    \changeref 
    \caption{\changeline{The performance of \method when integrated with RepoCoder and GraphCoder.}}
    \label{D2:complement}
    \begin{tabular}{lcccc}
        \toprule
        \multicolumn{5}{c}{\textbf{\method + RepoCoder}} \\
        \midrule
        \multirow{2}{*}{\textbf{Model}} & \multicolumn{2}{c}{\textbf{Oracle}} & \multicolumn{2}{c}{\textbf{Iteration-2}} \\
        \cmidrule(lr){2-3} \cmidrule(lr){4-5}
        & EM & BLEU & EM & BLEU \\
        \midrule
        aiXcoder-7B & 48.11 & 58.41 & 38.56 & 49.28 \\
        \rowcolor[rgb]{ .741,  .843,  .933} aiXcoder-7B-v2 & \textbf{53.94} & \textbf{64.77} & \textbf{45.06} & \textbf{56.76} \\
        \multicolumn{1}{c}{\textbf{Relative Improve.}} & +12.1\% & +10.9\% & +16.9\% & +15.2\% \\
        \midrule
        \multicolumn{5}{c}{\textbf{\method + GraphCoder}} \\
        \midrule
        \multirow{2}{*}{\textbf{Model}} & \multicolumn{2}{c}{\textbf{Line-level}} & \multicolumn{2}{c}{\textbf{API-level}} \\
        \cmidrule(lr){2-3} \cmidrule(lr){4-5}
        & EM & BLEU & EM & BLEU \\
        \midrule
        aiXcoder-7B & 10.50 & 16.57 & 6.72 & 15.06 \\
        \rowcolor[rgb]{ .741,  .843,  .933} aiXcoder-7B-v2 & \textbf{13.63} & \textbf{20.61} & \textbf{9.11} & \textbf{20.09} \\
        \multicolumn{1}{c}{\textbf{Relative Improve.}} & +29.7\% & +24.4\% & +35.6\% & +33.4\% \\
        \bottomrule
    \end{tabular}
    \vspace{-0.4cm}
\end{table}

A natural concern is whether \method improves repo-level code completion but hurts the performance on other tasks. To address this concern, we evaluate the performance of aiXcoder-7B and aiXcoder-7B-v2 on single-file completion and general code generation. For single-file completion, we remove all cross-file contexts in \method-132K and evaluate models on the new data. For code generation, we evaluate models on a popular code generation benchmark - HumanEval \cite{humaneval}. We report the average of EM and BLEU for single-file completion, and Pass@1 for code generation.

\textbf{Results and Analyses.} The results are shown in Table~\ref{D1:single_humaneval}. On single-file completion, the EM of aiXcoder-7B-v2 reaches 53.1, compared to 46.1 of aiXcoder-7B, with an improvement of 15.2\%. On HumanEval, the Pass@1 of aiXcoder-7B-v2 reaches 51.8, compared to 43.2 of aiXcoder-7B, with an improvement of 20.0\%. These results show that \method can improve the code generation and completion capabilities of LLMs beyond repo-level code completion.

\subsection{Complementary to Existing Code Completion Methods}
\label{sec:discussion:complement}

To validate the complementarity between our work and existing code completion approaches, we apply aiXcoder-7B and aiXcoder-7B-v2 (trained by our work) into two representative repo-level code completion approaches (\ie RepoCoder \cite{RepoCoder} and GraphCoder \cite{GraphCoder}). We conduct experiments on the original benchmarks in RepoCoder and GraphCoder. The results are shown in Table~\ref{D2:complement}. RepoCoder / GraphCoder with aiXcoder-7B-v2 outperforms that with aiXcoder-7B. For example, RepoCoder achieves an EM improvement from 38.56 to 45.06 after Iteration-2, while GraphCoder improves from 10.50 to 13.63 at the line level. These results indicate that our work complements existing repo-level code completion approaches.

\end{change}

\section{Threats to Validity}
\label{sec:threat}

There are four main threats to the validity of our work:

\textbf{Data leakage.} It means that the test data is leaked to the training data, resulting in unfair evaluations. Our experiments are conducted on two benchmarks, \ie \method-132K (Test set) and CrossCodeEval. We take three measures to address the data leakage threat. First, the training data of our studied LLMs is cut off as of March 2024. Thus, we only collect repositories created after March 2024 to construct \method-132K (Test set). Thus, the test data of \method-132K does not appear in the training data of LLMs. Second, when collecting \method-132K, we excluded repositories and the ground-truth code in CrossCodeEval. Thus, CrossCodeEval is not leaked to the training data of \method-132K. Third, as stated in the original paper of aiXcoder-7B \cite{aixcoder}, the developers of aiXcoder-7B have excluded CrossCodeEval from the pre-training data. Thus, the improvements of aiXcoder-7B-v2 on CrossCodeEval are not impacted by data leakage.

\textbf{Limited types of context-aware code.} Our \method-132K only considers two types of context-aware code (\ie cross-file API invocations and code spans similar to cross-file context). Our motivation is that these two types of code have explicit and common relationships with cross-file context (\eg calling, high code similarity). Thus, we can leverage parsers and code retrievers to collect these types of code for training LLMs. We also notice that there may be \textit{implicit} relationships between code and the cross-file context, for example, providing background of the current repository. However, it is hard to define implicit relationships and collect such type of code. Thus, we leave other types of context-aware to future work.

\textbf{The performance on generating code without similar code.} As stated in Section \ref{sec:CoLT:data:collection}, \method-132K only retains code spans whose maximum similarity with the cross-file context exceeds the threshold (\ie 0.3). Thus, one threat is: how well does the model perform when the code to be completed has low similarity to the cross-file context? To address this threat, we collect code spans whose maximum similarity with the cross-file context is lower than the threshold. 
aiXcoder-7B-v2 correctly completes 320 code spans, and aiXcoder-7B correctly completes 217 code spans,
demonstrating improved performance on low-similarity cases.

\textbf{The impact of repository versions.} Following standard practices  \cite{RepoCoder,CoCoMIC,CrossCodeEval,GraphCoder}, we extract the in-file and cross-file context from the current version of the repositories. However, when developers write the ground-truth code, the repository in a previous version, and the context may be slightly different. This problem is unexplored and out of the scope of our work. Besides, it is hard to precisely extract context due to the tangled commits. To address this threat, we select 10 repositories from \method-132K, manually add an incomplete function, and ask LLMs to complete the function. aiXcoder-7B-v2 correctly completes 6 functions, while aiXcoder-7B-v2 only completes 3 functions. The results avoid the impact of repository versions and show the effectiveness of \method.

\section{Related Work}
\label{sec:related_work}

\textbf{Repo-level Code Completion.} Repo-level code completion has emerged as a critical challenge in software engineering, aiming to enhance code completion accuracy by leveraging the broader repository context. Prior studies have predominantly focused on the retrieval of relevant context from the repository, with various strategies proposed to extract pertinent code snippets \cite{A3-CodGen, RLCoder, Dataflow, Monitor}. 
For instance, RepoCoder \cite{RepoCoder} introduces an iterative retrieval and generation approach.
Similarly, CrossCodeEval \cite{CrossCodeEval} proposes a straightforward retrieval method, using the lines preceding the cursor as a query to extract and append similar code snippets. GraphCoder \cite{GraphCoder} advances this by constructing a Code Context Graph (CCG) to represent the repository, enabling more sophisticated retrieval based on control flow, data flow, and dependencies. While these approaches have made significant strides in context retrieval, how to effectively utilize the retrieved context remains underexplored. Most existing studies view LLMs as a ``perfect'' generator and simply input long contexts into LLMs.
However, this simplistic approach may not fully exploit the potential of the retrieved information. To address this limitation, FT2Ra \cite{FT2Ra} integrates retrieval results into the generation process by modifying the model’s logits based on the retrieved content. Similarly, COCOMIC \cite{CoCoMIC} fuses retrieved cross-file contexts with in-file contexts within the model’s self-attention mechanism. However, both FT2Ra and COCOMIC require modifications to the model architecture, making them unsuitable for already pre-trained LLMs.

Our work diverges from this retrieval-centric paradigm by focusing on how to effectively assimilate and utilize the long cross-file context through a novel approach. In practice, our approach is complementary to the above related work.

\begin{change}
\textbf{Repo-level Pre-training.} Repo-level pre-training has become a popular approach to enhancing LLMs' long-context code completion capability \cite{LC_Train,CodeGemma,DeepSeek-Coder,Qwen2.5-Coder}, creating long-context samples by co-locating relevant files within repositories and continually pre-training LLMs on these samples.

Compared with existing repo-level pre-training studies, our work has the following advantages: (1) Data construction. Existing repo-level pre-training focuses on how to pack cross-file context, such as dependency graph-based packing and unit test-based lexical packing in CodeGemma \cite{CodeGemma}. However, the ground truths may be irrelevant to the cross-file context, making it difficult to effectively enhance long-context code completion. In contrast, our work emphasizes context-aware code completion. The ground truths (\eg cross-file API invocations) are highly relevant to the cross-file context, thus benefiting long-context code completion. (2) Training techniques. Existing studies mainly employ the next-token prediction training, whereas we introduce reinforcement learning to explicitly penalize LLMs when they fail to utilize the context effectively. 
(3) Further improvements. Models such as aiXcoder-7B and DeepSeek-Coder already perform repo-level pre-training. Our proposed approach can still further improve their performance, demonstrating the effectiveness and complementarity of our work.
\end{change}

\section{Conclusion and Future Work}
\label{sec:conclusion}

In this paper, we reveal a limitation of LLMs in repo-level code completion, \ie LLMs struggle to fully utilize information (\eg relevant APIs or similar code) in the cross-file context. To address this limitation, we propose \method, a purely data-driven approach to explicitly teach LLMs to focus on the cross-file context. \method constructs a large-scale repo-level code completion dataset - \method-132K and trains LLMs to generate context-aware code (\ie cross-file API invocations and code spans similar to cross-file context). We apply \method to multiple LLMs (\eg aiXcoder-7B) and conduct extensive experiments on \method-132K and CrossCodeEval. Results show that \method substantially improves the performance of multiple LLMs in repo-level code completion. 

In the future, we will further improve the ability of LLMs to process lengthy contexts. On the one hand, we plan to explore effective approaches to expanding the context window of LLMs. On the other hand, we will study new training techniques to help LLMs learn long-range dependencies.

\section*{Acknowledgment}

This research is supported by the National Key R\&D Program under Grant No. 2023YFB4503801, the National Natural Science Foundation of China under Grant No. 62192733, 62192730, 62192731, the Major Program (JD) of Hubei Province (No.2023BAA024), and the Beijing Natural Science Foundation under Grant No. 4264107.

\bibliographystyle{IEEEtran}
\bibliography{sample-base}

\end{document}